\documentclass[aps,prl,twocolumn,10pt,amsmath,amssymb,bibnotes,superscriptaddress,longbibliography]{revtex4-1}

\usepackage{graphicx}
\usepackage{siunitx}
\usepackage[colorlinks,urlcolor=blue,citecolor=blue,linkcolor=blue]{hyperref}

\newcommand{\ket}[1]{\ensuremath{\left|#1\right\rangle}}

\begin{document}

\title{Direct Entropy Measurement in a Mesoscopic Quantum System}
\author{Nikolaus Hartman}
\email{nik.hartman@gmail.com}
	\affiliation{Stewart Blusson Quantum Matter Institute, University of British Columbia, Vancouver, British Columbia, V6T1Z4, Canada}
	\affiliation{Department of Physics and Astronomy, University of British Columbia, Vancouver, British Columbia, V6T1Z1, Canada}
\author{Christian Olsen}
	\affiliation{Stewart Blusson Quantum Matter Institute, University of British Columbia, Vancouver, British Columbia, V6T1Z4, Canada}
	\affiliation{Department of Physics and Astronomy, University of British Columbia, Vancouver, British Columbia, V6T1Z1, Canada}
\author{Silvia L\"{u}scher}
	\affiliation{Stewart Blusson Quantum Matter Institute, University of British Columbia, Vancouver, British Columbia, V6T1Z4, Canada}
	\affiliation{Department of Physics and Astronomy, University of British Columbia, Vancouver, British Columbia, V6T1Z1, Canada}
\author{Mohammad Samani}
	\altaffiliation{The Hospital for Sick Children, Toronto, ON, Canada}
	\altaffiliation{Fields Institute for Research in Mathematical Sciences, Toronto, ON, Canada}
	\affiliation{Stewart Blusson Quantum Matter Institute, University of British Columbia, Vancouver, British Columbia, V6T1Z4, Canada}
	\affiliation{Department of Physics and Astronomy, University of British Columbia, Vancouver, British Columbia, V6T1Z1, Canada}
\author{Saeed Fallahi}
	\affiliation{Department of Physics and Astronomy, Purdue University, West Lafayette, Indiana, USA}
	\affiliation{Station Q Purdue, Purdue University, West Lafayette, Indiana, USA}
	\affiliation{Birck Nanotechnology Center, Purdue University, West Lafayette, Indiana, USA}
\author{Geoffrey C. Gardner}
	\affiliation{Station Q Purdue, Purdue University, West Lafayette, Indiana, USA}
	\affiliation{Birck Nanotechnology Center, Purdue University, West Lafayette, Indiana, USA}
    	\affiliation{School of Materials Engineering, Purdue University, West Lafayette, Indiana, USA}
\author{Michael Manfra}
	\affiliation{Department of Physics and Astronomy, Purdue University, West Lafayette, Indiana, USA}
	\affiliation{Station Q Purdue, Purdue University, West Lafayette, Indiana, USA}
	\affiliation{Birck Nanotechnology Center, Purdue University, West Lafayette, Indiana, USA}
	\affiliation{School of Electrical and Computer Engineering,  Purdue University, West Lafayette, Indiana, USA}
    	\affiliation{School of Materials Engineering, Purdue University, West Lafayette, Indiana, USA}
\author{Joshua Folk}
\email{jfolk@physics.ubc.ca}
	\affiliation{Stewart Blusson Quantum Matter Institute, University of British Columbia, Vancouver, British Columbia, V6T1Z4, Canada}
	\affiliation{Department of Physics and Astronomy, University of British Columbia, Vancouver, British Columbia, V6T1Z1, Canada}
\date{\today}

\maketitle

\textbf{The entropy of an electronic system offers important insights into the nature of its quantum mechanical ground state. This is particularly valuable in cases where the state is difficult to identify by conventional experimental probes, such as conductance. Traditionally, entropy measurements are based on bulk properties, such as heat capacity, that are easily observed in macroscopic samples but are unmeasurably small in systems that consist of only a few particles \cite{Ramirez1999,Schmidt2017}. In this work, we develop a mesoscopic circuit to directly measure the entropy of just a few electrons, and demonstrate its efficacy using the well understood spin statistics of the first, second, and third electron ground states in a GaAs quantum dot \cite{Tarucha1996, Ciorga2000, Duncan2000, Lindemann2002, Potok2003, Hofmann2016}.  The precision of this technique, quantifying the entropy of a single spin-$\frac{1}{2}$ to within 5\% of the expected value of $k_B \ln{2}$, shows its potential for probing more exotic systems.  For example, entangled states or those with non-Abelian statistics could be clearly distinguished by their low-temperature entropy\cite{Cooper2009, Ben-Shach2013, Smirnov2015, Hou2012, Alkurtass2016}.}

\begin{figure}
        \includegraphics[width=1.0\columnwidth]{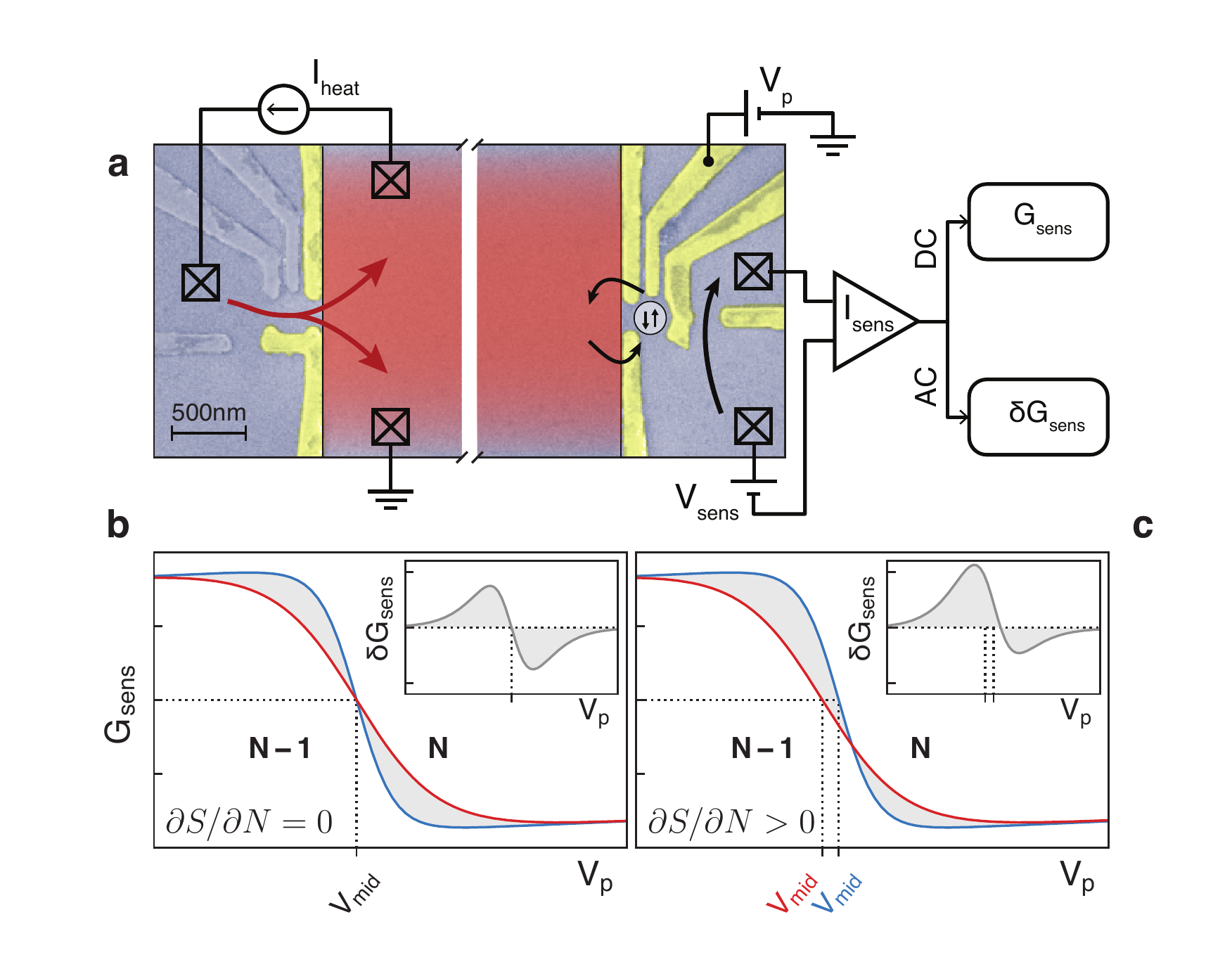}
        \caption{\label{fig:fig1} \textbf{Measurement protocol}  (a) Scanning electron micrograph of a device similar to the one measured. Electrostatic gates (gold) define the circuit in a 2D electron gas (2DEG), with grey gates grounded. Squares indicate ohmic contacts to the 2DEG.  The temperature of the electron reservoir in the middle (red) is oscillated using AC current, $I_{heat}$, at frequency $f_{heat}$ through the quantum point contact (QPC) on the left.  A portion of the $5~\mu$m-wide reservoir has been removed here for clarity.  The occupation of the quantum dot, tunnel coupled to the right side the reservoir, is tuned by $V_p$ and monitored by $I_{sens}$ through the charge sensor QPC. $I_{sens}$ is split into DC and AC components, the latter being measured by a lock-in amplifier at $2f_{heat}$.  (b) and (c) Simulated DC charge sensor signal, $G_{sens}$, for a transition from $N-1 \rightarrow N$ electrons at two temperatures ($T_{Red} > T_{Blue}$), showing two possible cases for $\frac{\partial S}{\partial N}$. Insets show the corresponding difference, $\delta G_{sens}$, between hot and cold curves.}
\end{figure}

Our approach is analogous to the milestone of spin-to-charge conversion achieved over a decade ago, in which the infinitesimal magnetic moments of a single spin were detected by transforming them into the presence or absence of an electron charge \cite{Elzerman2004, Ono2004}.  Following this example, we perform an entropy-to-charge conversion, making use of the Maxwell relation
\begin{align}
\label{eqn:max}
        \left(\frac{\partial \mu}{\partial T}\right)_{p,N} &= -\left(\frac{\partial S}{\partial N}\right)_{p,T}
\end{align}
that connects changes in entropy, particle number, and temperature ($S$, $N$, and $T$, respectively) to changes in the chemical potential, $\mu$, a quantity that is simple to measure and control.

The Maxwell relation in Eq.~\ref{eqn:max} forms the basis of two theoretical proposals to measure non-Abelian exchange of Moore-Read quasiparticles in  the $\nu = \frac{5}{2}$ state via their entropy \cite{Cooper2009,Ben-Shach2013}.  Reference~\onlinecite{Ben-Shach2013} proposes a strategy by which quasiparticle entropy could be deduced from the temperature-dependent shift of charging events on a local disorder potential---a thermodynamic equivalent of the measurements that established the $e/4$ quasiparticle charge\cite{Venkatachalam2011}. As a demonstration of the viability and the high accuracy achievable by this technique, we investigate a well-understood system with localized fermions in place of more exotic quasiparticles: a few-electron GaAs quantum dot. The entropies of the first three electron states in the dot are measured by the temperature-dependent charging scheme laid out in Ref.~\onlinecite{Ben-Shach2013}.   Applying the language of quantum dots to Eq.~\ref{eqn:max}, the entropy difference between the $N-1$ and $N$ electron ground states ($\Delta S_{N-1\rightarrow N}$ for $\Delta N=1$) is measured via the shift with temperature in the electrochemical potential, $\mu_N$, needed to add the $N$th electron to the dot.

The measurement relies on the mesoscopic circuit shown in Fig.~\ref{fig:fig1}a, using electrostatic gates to realize an electron reservoir in thermal and diffusive equilibrium with a few-electron quantum dot coupled to its right side.  The occupation of the dot is tuned with the plunger gate voltage, $V_p$, and measured using an adjacent quantum point contact as a charge sensor \cite{Field1993, Staring2007, Thierschmann2015}.  Applying more positive $V_p$ lowers $\mu_N$,  bringing the $N$th electron into the dot when $\mu_{N}$ drops below the Fermi level of the reservoir, $E_F$. The reservoir temperature, $T$, can be increased above the GaAs substrate temperature by Joule heating from current, $I_{heat}$, driven through a quantum point contact on the left side.   Charge transitions on the dot appear as steps in the charge sensor conductance, $G_{sens}(V_p)$, thermally broadened by the reservoir temperature (Figs.~\ref{fig:fig1}b and c).  The gate voltage corresponding to the midpoint of the transition, $V_{mid}$, marks the electrochemical potential at which the probabilities of finding $N-1$ and $N$ electrons on the dot are equal.

When $\mu_N$ shifts with temperature, $V_{mid}$ also shifts; it is the shift in $V_{mid}$ with temperature that forms the basis of our experiment (Fig.~\ref{fig:fig1}c).  In practice, charge noise limits the accuracy to which $V_{mid}$ can be measured. To overcome this, the measurement is done with a lock-in amplifier, oscillating the temperature using an AC $I_{heat}$ and measuring resultant oscillations in $G_{sens}$, which we label $\delta G_{sens}$.  As seen in the insets of Figs.~\ref{fig:fig1}b and c, the lineshape of $\delta G_{sens}$ is perfectly antisymmetric when $\partial S/\partial N=0$, but asymmetric when $\partial S/\partial N \neq 0$.

The temperature-induced shift in the dot chemical potential with respect to reservoir $E_F$ can also be understood in terms of detailed balance.  At $V_{mid}$, where probabilities for $N$ and $N-1$ electrons on the dot are equal, the tunnel rates $\Gamma_{in}=\Gamma_{N-1\rightarrow N}$ and $\Gamma_{out}=\Gamma_{N\rightarrow N-1}$ must also be equal. These rates depend on the number of available states in the tunneling process, and therefore on the degeneracies, $d_{N-1}$ and $d_{N}$, of the $N-1$ and $N$ ground states \cite{Beenakker1991, Gustavsson2009}.  The condition $\Gamma_{in} = \Gamma_{out}$ leads to a simple relationship between degeneracy and the thermally broadened Fermi function, $f(\mu_{N}-E_{F}, T)$: $d_{N-1}/d_{N}=f/(1-f)$. Using the Boltzman entropy, $S_{N}=k_{B} \ln{d_N}$, this relationship becomes $\Delta S_{N-1\rightarrow N}= (\mu_{N}-E_F)/T$, clearly demonstrating the connection between entropy, temperature, and the shift in $\mu_N$ at $V_{mid}$. Previous experiments have explored the relationship between tunnel rates and degeneracy using time-resolved transport spectroscopy and by coupling quantum dots to atomic force cantilever oscillations \cite{Cockins2010, Bennett2010, Beckel2014, Hofmann2016}. The approach presented here is a thermodynamic analogue, and extends entropy measurements to a wider set of applications where tunneling processes may not be observable in real-time.

\begin{figure}
        \includegraphics[width=1.0\columnwidth]{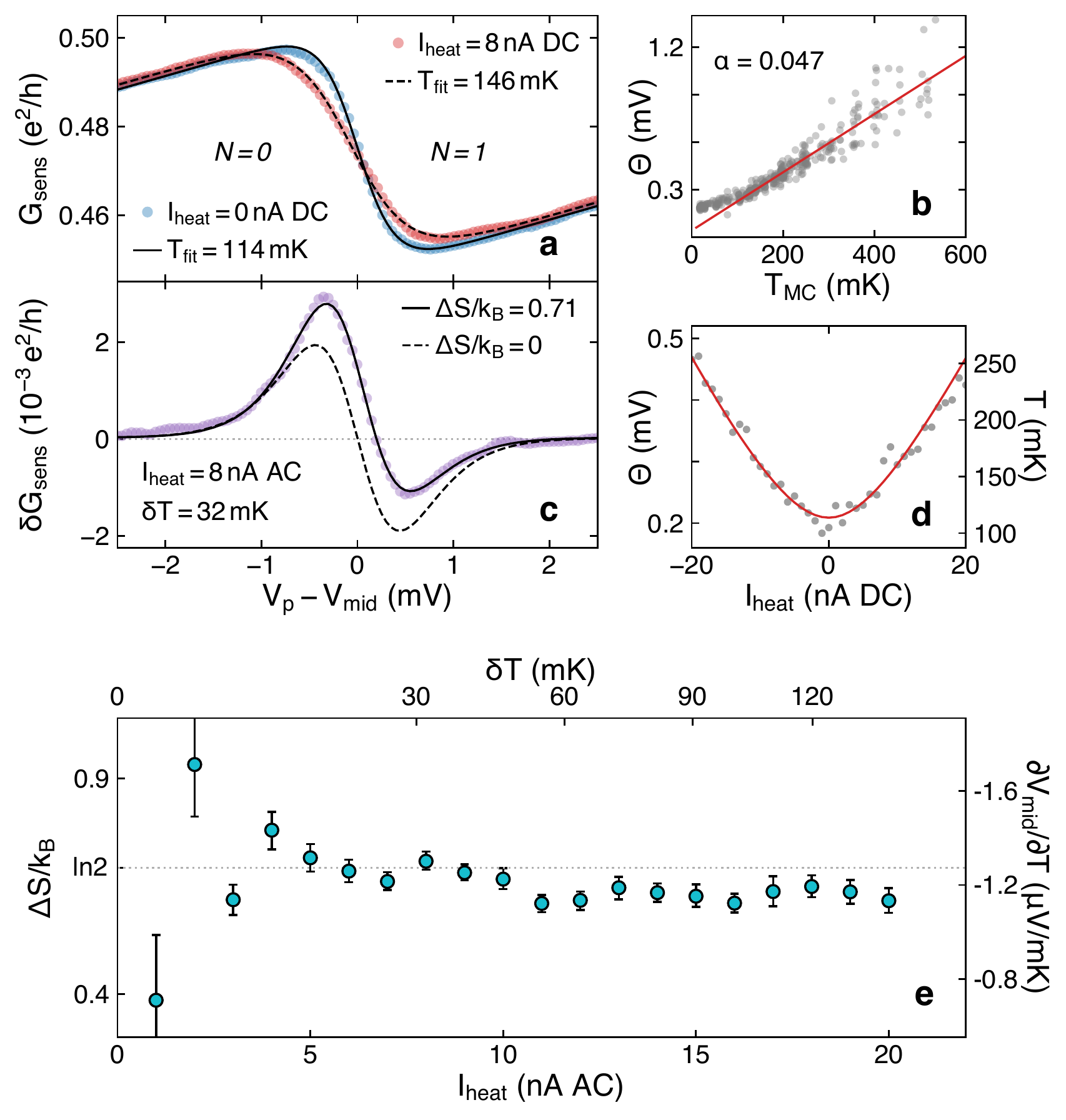}
        \caption{\label{fig:fig2} \textbf{Entropy measurement for a single spin-$\mathbf{\frac{1}{2}}$} (a) Charge sensor data for $N=0 \rightarrow 1$ at two temperatures set by DC current through the QPC heater. (b) Transition width, $\Theta$, was linear in $T_{MC}$ above 100 mK, for $I_{heat}=0$. Lever arm $\alpha$ is calculated by fitting a straight line to this region.  (c) Lock-in measurement of $\delta G_{sens}$ with $\delta T = \SI{32}{\milli\kelvin}$, determined from the calibration in panel (d). Fits to $\delta G_{sens}$ (Eq.~\ref{eqn:dg-sens}) are shown with $\Delta S / k_B$  as a free parameter (solid) and fixed at $\Delta S / k_B=0$ (dashed). (d) $\Theta$ grows with DC current through the QPC heater. A fit to $T^2 = aT_{MC}^2+b I^2_{heat}R_{QPC}$ is used to convert between $I_{heat}$ and $\delta T$, where $T_{MC}$ is the mixing chamber temperature\cite{Mittal1996}. (e)  Entropy measurements were independent of the magnitude of $I_{heat}$ oscillations over a large range. The top axis indicates the corresponding magnitude of $\delta T$, while the right axis shows the entropy signal converted to a gate voltage shift per unit temperature. Error bars show 95\% confidence intervals calculated with the bootstrap method.}
\end{figure}

The dot was tuned such that the source was weakly tunnel-coupled to the reservoir with the drain closed.  The conductance of the charge sensor was tuned to $G_{sens}{\sim}e^2/h$, where it was most sensitive to charge on the dot.  The addition of the first electron to the dot was marked by a decrease in $G_{sens}$ that is consistent with a thermally-broadened two-level transition (Fig.~\ref{fig:fig2}a):
\begin{align}
\label{eqn:g-sens}
        G_{sens}(V_p,\Theta) &= G_0 \tanh\left(\frac{V_p - V_{mid}(\Theta)}{2\Theta}\right)  \\
                        &\quad + \gamma_1 V_{p} + G_2 \nonumber
\end{align}
where $G_0$ quantifies the sensor sensitivity, $\Theta = \frac{k_B T}{\alpha e}$ is the thermal broadening expressed in units of gate voltage, $\alpha\equiv \frac{1}{e}\frac{d \mu_{N}}{d V_p}$ is the lever arm,  $\gamma_1$ reflects the cross capacitance between the charge sensor and plunger gate, and $G_2$ is an offset. Figure~\ref{fig:fig2}a shows two such transition curves with thermal broadening set by $I_{heat}$. For $I_{heat}=0$, $\Theta$ followed $T_{MC}$ down to approximately \SI{100}{\milli\kelvin} (Fig.~\ref{fig:fig2}b), validating the approximation of thermal broadening used throughout this experiment.

The data in Fig.~\ref{fig:fig2}c, and corresponding fits, illustrate a measurement of $\Delta S_{0\rightarrow 1}$ across the $0 \rightarrow 1$ electron transition. The lock-in measurement of $\delta G_{sens}$, due to temperature oscillations $\delta T$, yields the characteristic peak-dip structure seen in Fig.~\ref{fig:fig2}c.

The expected lineshape of such a curve is $\delta G_{sens} = \frac{\partial G_{sens}}{\partial T} \delta T$, with $G_{sens}$ defined by Eq. \ref{eqn:g-sens}.  This lineshape depends explicitly on $\Delta S$, recognizing (via Eq.~\ref{eqn:max}) that $\frac{\partial V_{mid}}{\partial \Theta}=\frac{1}{k_B}\frac{\partial \mu}{\partial T} =-\frac{1}{k_B}\Delta S_{N-1\rightarrow N}$:
\begin{align}
\label{eqn:dg-sens}
        \delta G_{sens}(V_p, \Theta) &\propto -\delta T \left[ \frac{V_p - V_{mid}(\Theta)}{2\Theta} - \frac{\Delta S}{2k_B} \right]\times \\
        				      &\quad\cosh^{-2}\left(\frac{V_p - V_{mid}(\Theta)}{2\Theta}\right) + const. \nonumber
\end{align}
\noindent As expected  from Figs.~\ref{fig:fig1}b and c, $\delta G_{sens}(V_p)$ is antisymmetric around $V_{mid}$ for $\Delta S=0$, and asymmetric for $\Delta S\neq 0$.  A fit of the data in Fig.~\ref{fig:fig2}c to Eq.~\ref{eqn:dg-sens} yields $\Delta S_{0\rightarrow 1}=(1.02 \pm 0.03) k_B \ln{2}$, closely matching the expected $\Delta S_{0 \rightarrow 1} = S_1 - S_0 =k_B\ln{2}$ for transitions between an empty dot with zero entropy ($S_0=0$) and the two-fold degenerate one-electron state ($d_1=2$) with entropy $S_1=k_B \ln{2}$.

It is important to note that $\Delta S$ is extracted from fits to Eq.~\ref{eqn:dg-sens} based solely on the asymmetry of the lineshape, with no calibration of measurement parameters (such as $\delta T$ or the lever arm $\alpha$) required.  We can, however, estimate $\alpha$ and $\delta T$ by determining $\Theta$ from fits to Eq.~\ref{eqn:g-sens} for varying substrate temperature (Fig.~\ref{fig:fig2}b) and $I_{heat}$ (Fig.~\ref{fig:fig2}d). Measurements of $\Delta S$ remained constant over a broad range of $\delta T$ (Fig.~\ref{fig:fig2}e), as expected for temperatures low enough not to excite orbital degrees of freedom on the dot.

\begin{figure}
        \includegraphics[width=1.0\columnwidth]{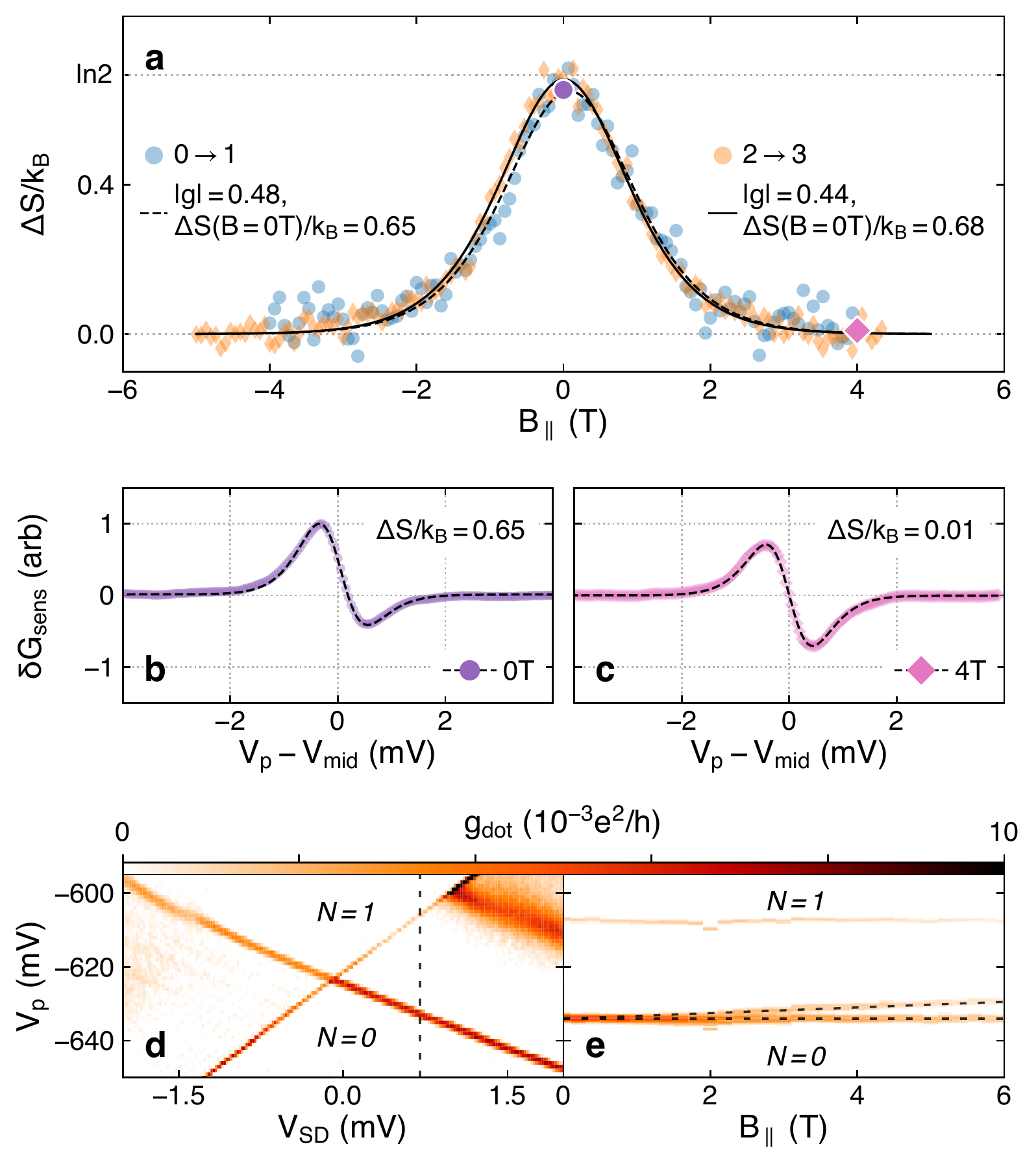}
        \caption{\label{fig:fig3} \textbf{Magnetic field dependence} (a) Changes in entropy for $N=0 \rightarrow 1$ and $2 \rightarrow 3$ transitions, overlaid to highlight similar behaviors.  Each data point corresponds to a single $\delta G_{sens}(V_p)$ fit; multiple scans are carried out at various in-plane magnetic fields.   (b) and (c) Characteristic $\delta G_{sens}$ traces from which the data in (a) were extracted. The two data points corresponding to (b) and (c) are shown as large markers in (a). (d) Bias spectroscopy data for the $N=0 \rightarrow 1$ transition. Dashed line at $V_{SD}$ = \SI{700}{\micro\electronvolt} shows where data in (e) are taken. (e) Fixed bias data showing fits to Zeeman splitting of the ground state (dashed lines) from which $|g| = 0.42 \pm 0.01$ is extracted.}
\end{figure}

Confirmation that the measured $\Delta S$ derives from spin degeneracy is seen through its evolution with in-plane magnetic field, $B_\parallel$. Figure \ref{fig:fig3}a compares $\Delta S(B_\parallel)$ for the $0 \rightarrow1 $ and $2 \rightarrow 3$ transitions, both of which correspond to transitions from total spin zero to total spin one-half. The entropies of the one- and three-electron states go to zero as Zeeman splitting lifts the spin degeneracy, following the Gibbs entropy for a two-level system:
\begin{align}
\label{eqn:gibbs}
        S &= k_B \sum_{i=\pm} p_{i}(B_\parallel, T) \ln{ p_{i}(B_\parallel,T) }
\end{align}
where $p_{\pm}(B_\parallel, T) = (1+ e^{\mp \frac{g\mu_B B_{\parallel}}{k_B T}})^{-1}$ are the probabilities for the unpaired electron to be in the spin up or spin down states at a given field and temperature. Fits to Eq.~\ref{eqn:gibbs}, with the ratio $g/T$ and an added scaling $\Delta S(B=0)$ as free parameters, give $\Delta S_{0 \rightarrow 1}(B=0)=(0.94 \pm 0.03) k_B \ln{2}$ and $\Delta S_{2 \rightarrow 3}(B=0)=(0.98 \pm 0.02) k_B \ln{2}$ (Fig.~\ref{fig:fig3}), and reflect the collapse to zero at high field where spin degeneracy is broken. This collapse can also be seen qualitatively, in the crossover from asymmetric to antisymmetric lineshapes of $\delta G_{sens}(V_p)$ (Figs.~\ref{fig:fig3}b and c). Estimating an average $T$ for each data set using the calibration in Fig.~\ref{fig:fig2}d yields $|g|=0.48 \pm 0.02$ and $|g|=0.44 \pm 0.01 $ for the $0\rightarrow 1$ and $2\rightarrow 3$ transitions, respectively. Errors in the g-factor measurement are likely due to the difficulty of estimating temperature oscillations. Still, the g-factors are consistent with reported values\cite{Cronenwett1998,Hanson2003,Zumbuhl2004} and the value measured separately in Fig.~\ref{fig:fig3}e using bias spectroscopy.

\begin{figure}
        \includegraphics[width=1.0\columnwidth]{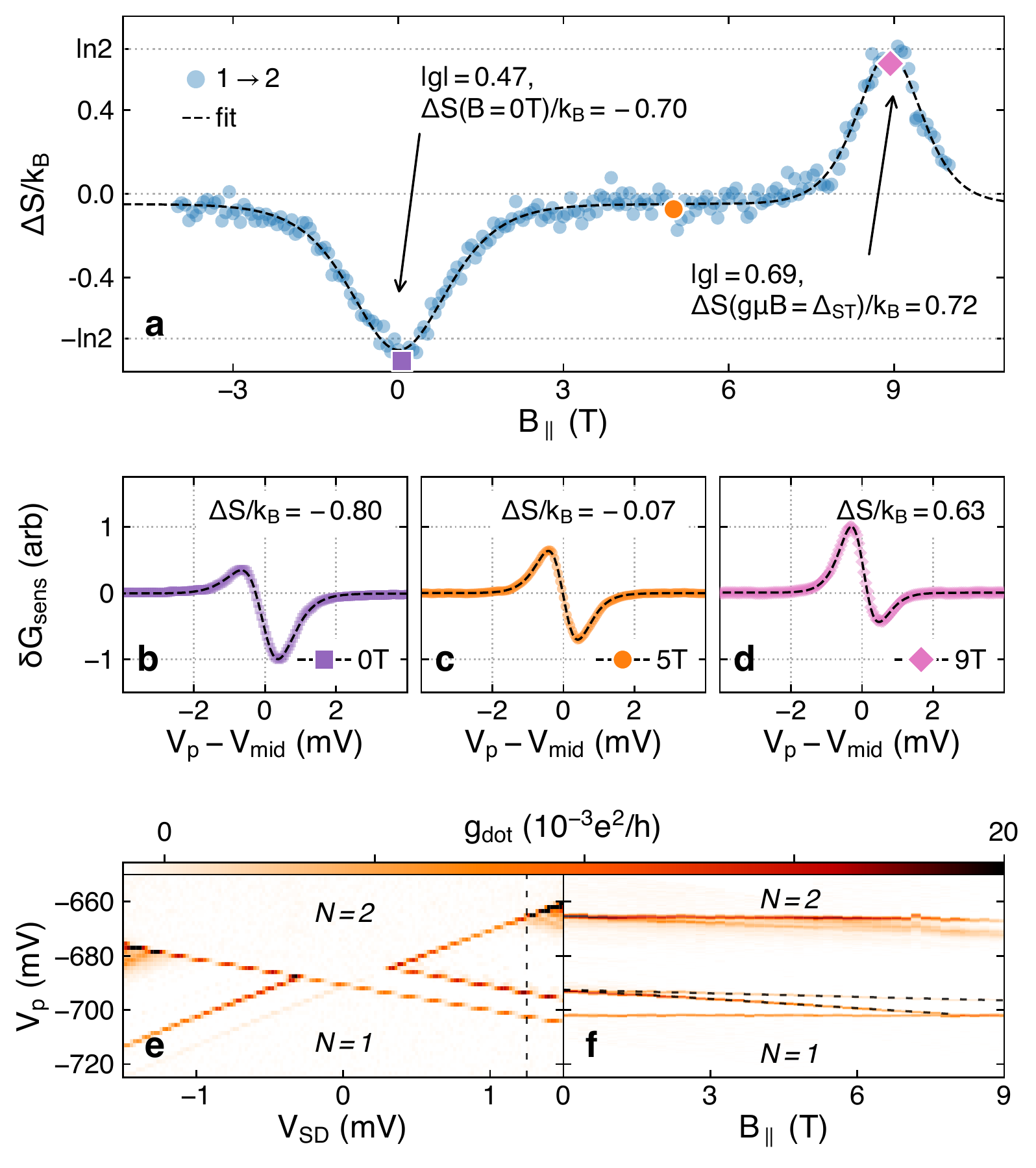}
        \caption{\label{fig:fig4}  \textbf{Entropic signature of a singlet-triplet crossing} (a) Change in entropy, extracted from $\delta G_{sens}$ fits at varying in-plane field. Dashed line shows fit to Eq.~\ref{eqn:gibbs}, allowing for an offset from $\Delta S=0$ away from the degenerate points to compensate for non-linearities in the charge sensor. Values stated for $\Delta S$ are with respect to the vertical offset apparent in the data. (b), (c), and (d) show characteristic $\delta G_{sens}$ traces from which the data in (a) were extracted. These data points are show as large markers in (a). (e) Bias spectroscopy data for the $N=1 \rightarrow 2$ transition. Transitions to the two-electron triplet state correspond to the lines appearing at $V_{SD} = \SI{\pm 320}{\micro\electronvolt}$. Dashed line at $V_{SD}$ = \SI{1250}{\micro\electronvolt} shows where data in (f) are taken. (f) Fixed bias data in in-plane field. Triplet level is split into $\ket{\mathcal{T}_+}$ and $\ket{\mathcal{T}_0}$ levels with a third $\ket{\mathcal{T}_-}$ level not visible here. At \SI{8.4}{\tesla} $\ket{\mathcal{T}_+}$ becomes degenerate with $\ket{\mathcal{S}}$. $|g|=0.40\pm0.04$ is determined using $\ket{\mathcal{T}_0}$ and $\ket{\mathcal{T}_+}$ fits (dashed).}
\end{figure}

The $1\rightarrow 2$ transition can be understood as the inverse of the $0 \rightarrow 1$ transition for $B_\parallel < \SI{5}{\tesla}$, comparing Figs.~\ref{fig:fig3}a and \ref{fig:fig4}a. For relatively low fields, the two-electron ground state remains a spin singlet with zero entropy, while the one-electron entropy goes from $k_B\ln{2}$ to zero due to Zeeman splitting.  At higher fields, the one-electron ground state remains non-degenerate while the two-electron ground state gains a two-fold degeneracy when the singlet $\ket{\mathcal{S}}$ and triplet \ket{\mathcal{T}_+} states cross.  This singlet-triplet crossing is seen in bias spectroscopy data (Fig.~\ref{fig:fig4}f) at \SI{8.4}{\tesla}, and in the appearance of a peak in $\Delta S_{1\rightarrow 2}$ at \SI{9}{\tesla} (Fig.~\ref{fig:fig4}a). The discrepancy in field required to drive the singlet-triplet degeneracy in Figs.~\ref{fig:fig4}a and f is attributed to a change in shape of the dot potential, caused by altering the confinment gate voltages, when transitioning from one to two open tunnel barriers.

The field-dependent entropy measurement for the $1 \rightarrow 2$ transition can again be fit using Eq.~\ref{eqn:gibbs}, with probabilities as before for the one-electron states and $p_{\ket{\mathcal{S}}}(B_\parallel, T) = (1+ e^{-\frac{g\mu_B B_\parallel - \Delta_\mathcal{ST}}{k_B T}})^{-1}$, $p_{\ket{\mathcal{T}_+}}(B_\parallel, T) = (1+ e^{+\frac{g\mu_B B_\parallel - \Delta_\mathcal{ST}}{k_B T}})^{-1}$ for the two-electron states, where $\Delta_\mathcal{ST}$ is the singlet-triplet splitting at zero field. From the fit, we find $\Delta S_{1\rightarrow 2}$ at the two-fold degenerate points, $B=0$ and \SI{9}{\tesla}, are $-(1.01\pm0.03) k_B \ln{2}$ and $(1.04 \pm 0.04) k_B \ln{2}$, respectively. The extracted g-factor, $|g| = 0.47 \pm 0.02$, from the peak at $B=0$ is consistent with the $0\rightarrow 1$ transition.  At the high-field singlet-triplet degeneracy we find $|g| = 0.69 \pm 0.04$, an unexpectedly high g-factor that is explained by a shift of the $\ket{\mathcal{T}_0}$ state with magnetic field, as seen in Fig.~\ref{fig:fig4}f and previous work \cite{Szafran2004}.

We conclude with a few notes to encourage the application of this entropy measurement protocol to other mesoscopic systems.  The crucial ingredients in achieving the high accuracy reported here were $i$) the ability to oscillate temperature rapidly enough to avoid $1/f$ noise, $ii$) the ability to measure charging transitions without perturbing the localized states, and $iii$) the fact that the charging transitions were thermally broadened.  Criterion $iii$) enabled the entropy determination purely by asymmetry, without the need to know $\delta T$ or other measurement parameters accurately, yielding an uncertainty less than 5\%.  With this level of precision, it should be possible, for example, to distinguish the $\frac{1}{2} k_B \ln{2}$ entropy of a non-Abelian Majorana bound state from the $k_B \ln{2}$ entropy of an Andreev bound state at an accidental degeneracy\cite{Smirnov2015, Hou2012}.   Similarly, the $S=\frac{1}{2} k_B \ln{2}$ two-channel Kondo state could be clearly distinguished from fully screened ($S=0$) or unscreened ($S=k_B \ln{2}$) spin states\cite{Alkurtass2016}.

\textbf{Methods} The device was built on a AlGaAs/GaAs heterostructure, hosting a 2D electron gas with density and mobility at \SI{300}{\milli\kelvin} of \SI{2.42e11}{\per\square\centi\metre} and \SI[per-mode=symbol]{2.56e6}{\square\centi\metre\per\volt\per\second} respectively, determined in a separate measurement.   Mesas and NiAuGe ohmic contacts to the 2DEG were defined by standard photolithography techniques, followed by atomic layer deposition of \SI{10}{\nano\metre} $\mathrm{HfO_2}$ to improve the gating stability in the device. Fine gate structures, shown in Fig.~\ref{fig:fig1}a, were defined by electron beam lithography and deposition of \SI{3/18}{\nano\metre} Ti/Au.

The measurement was carried out in a dilution refrigerator with a two-axis magnet. The 2DEG was aligned parallel to the main axis with the second axis used to compensate for sample misalignment. In practice, out-of-plane fields up to \SI{100}{\milli\tesla} showed no effect on our data. A retuning of the quantum dot gates was necessary to capture the bias spectroscopy data in Figs.~\ref{fig:fig3}d,e and \ref{fig:fig4}e,f. The rightmost gate (Fig.~\ref{fig:fig1}a) on the quantum dot was used to tune between the one and two lead configurations, for the entropy and bias spectroscopy measurements respectively. This tuning had a significant effect on the shape of the potential well, accounting for variations in parameters such as $g$ and $\Delta_{ST}$ between the two measurement configurations.  Charge sensor conductance was measured using a DC voltage bias of \SIrange{200}{350}{\micro\volt}; we find that Joule heating through the sensor does not affect our reservoir temperatures up to $V_{sens}\sim\SI{500}{\micro\volt}$. The DC current ($I_{sens}$) was measured using an analog-digital convertor while AC current ($\delta I_{sens}$) was measured using a lock-in amplifier. The DC conductance reported here is $G_{sens}=I_{sens}/V_{sens}$ while the oscillations are defined as $\delta G_{sens}=(\delta I_{sens})/V_{sens}$.

The temperature of the reservoir was raised above the substrate temperature using $I_{heat}$ at AC or DC, with the QPC heater set by gate voltages to \SI{20}{\kilo\ohm}. Applying AC current at $f_{heat} =$ \SI{48.7}{\hertz} yields an oscillating Joule power, $P_{heat} = I^2_{heat}R_{QPC}$. To leading order this gives oscillations in temperature, and therefore $\delta G_{sens}$, at $2f_{heat}$.  These are  captured by the lock-in amplifier at the second harmonic of $I_{heat}$.  Except where noted, measurements of $\Delta S$ were made at $\delta T \sim $ \SI{50}{\milli\kelvin}, although the error bars in Fig.~\ref{fig:fig2} demonstrate that the measurements would have been just as accurate with $\delta T$ set to \SI{30}{\milli\kelvin}. The fixed pressure condition of Eq.~\ref{eqn:max} is met by working well below the Fermi temperature of the 2DEG, $T_F \sim \SI{100}{\kelvin}$, where degeneracy pressure dominates \cite{Landau1980}.

\textbf{Data Availability} Data generated for, and analyzed in, this study are available at \url{https://github.com/nikhartman/spin_entropy}. The repository also contains all code necessary to complete the analysis and create each of the figures in this manuscript.

\textbf{Acknowledgements} The authors acknowledge John Martinis for a helpful discussion on the interpretation of our measurements.   NH, CO, SL, MS, and JF supported by the Canada Foundation for Innovation, the National Science and Engineering Research Council, CIFAR, and SBQMI.  SF, GG, and MM supported by the U.S. DOE Office of Basic Energy Sciences, Division of Materials Sciences and Engineering Award No. DE-SC0006671, with additional support from Nokia Bell Laboratories for the MBE facility gratefully acknowledged.

\textbf{Author Contributions} NH and CO fabricated the mesoscopic device. GaAs heterostructures and their characterization were provided by SF, GG, and MM. SL and MS worked on early versions of the experiment and provided helpful discussion. NH performed measurements and analyzed data. Manuscript written by NH and JF with additional feedback from all authors.

\bibliography{qdentropy}{}

\begin{thebibliography}{30}%
\makeatletter
\providecommand \@ifxundefined [1]{%
 \@ifx{#1\undefined}
}%
\providecommand \@ifnum [1]{%
 \ifnum #1\expandafter \@firstoftwo
 \else \expandafter \@secondoftwo
 \fi
}%
\providecommand \@ifx [1]{%
 \ifx #1\expandafter \@firstoftwo
 \else \expandafter \@secondoftwo
 \fi
}%
\providecommand \natexlab [1]{#1}%
\providecommand \enquote  [1]{``#1''}%
\providecommand \bibnamefont  [1]{#1}%
\providecommand \bibfnamefont [1]{#1}%
\providecommand \citenamefont [1]{#1}%
\providecommand \href@noop [0]{\@secondoftwo}%
\providecommand \href [0]{\begingroup \@sanitize@url \@href}%
\providecommand \@href[1]{\@@startlink{#1}\@@href}%
\providecommand \@@href[1]{\endgroup#1\@@endlink}%
\providecommand \@sanitize@url [0]{\catcode `\\12\catcode `\$12\catcode
  `\&12\catcode `\#12\catcode `\^12\catcode `\_12\catcode `\%12\relax}%
\providecommand \@@startlink[1]{}%
\providecommand \@@endlink[0]{}%
\providecommand \url  [0]{\begingroup\@sanitize@url \@url }%
\providecommand \@url [1]{\endgroup\@href {#1}{\urlprefix }}%
\providecommand \urlprefix  [0]{URL }%
\providecommand \Eprint [0]{\href }%
\providecommand \doibase [0]{http://dx.doi.org/}%
\providecommand \selectlanguage [0]{\@gobble}%
\providecommand \bibinfo  [0]{\@secondoftwo}%
\providecommand \bibfield  [0]{\@secondoftwo}%
\providecommand \translation [1]{[#1]}%
\providecommand \BibitemOpen [0]{}%
\providecommand \bibitemStop [0]{}%
\providecommand \bibitemNoStop [0]{.\EOS\space}%
\providecommand \EOS [0]{\spacefactor3000\relax}%
\providecommand \BibitemShut  [1]{\csname bibitem#1\endcsname}%
\let\auto@bib@innerbib\@empty
\bibitem [{\citenamefont {Ramirez}\ \emph {et~al.}(1999)\citenamefont
  {Ramirez}, \citenamefont {Hayashi}, \citenamefont {Cava}, \citenamefont
  {Siddharthan},\ and\ \citenamefont {Shastry}}]{Ramirez1999}%
  \BibitemOpen
  \bibfield  {author} {\bibinfo {author} {\bibfnamefont {A~P}\ \bibnamefont
  {Ramirez}}, \bibinfo {author} {\bibfnamefont {A}~\bibnamefont {Hayashi}},
  \bibinfo {author} {\bibfnamefont {R~J}\ \bibnamefont {Cava}}, \bibinfo
  {author} {\bibfnamefont {R}~\bibnamefont {Siddharthan}}, \ and\ \bibinfo
  {author} {\bibfnamefont {B~S}\ \bibnamefont {Shastry}},\ }\bibfield  {title}
  {\enquote {\bibinfo {title} {{Zero-point entropy in ‘spin ice'}},}\ }\href
  {http://dx.doi.org/10.1038/20619 http://10.0.4.14/20619} {\bibfield
  {journal} {\bibinfo  {journal} {Nature}\ }\textbf {\bibinfo {volume} {399}},\
  \bibinfo {pages} {333} (\bibinfo {year} {1999})}\BibitemShut {NoStop}%
\bibitem [{\citenamefont {Schmidt}\ \emph {et~al.}(2017)\citenamefont
  {Schmidt}, \citenamefont {Bennaceur}, \citenamefont {Gaucher}, \citenamefont
  {Gervais}, \citenamefont {Pfeiffer},\ and\ \citenamefont
  {West}}]{Schmidt2017}%
  \BibitemOpen
  \bibfield  {author} {\bibinfo {author} {\bibfnamefont {B.~A.}\ \bibnamefont
  {Schmidt}}, \bibinfo {author} {\bibfnamefont {K.}~\bibnamefont {Bennaceur}},
  \bibinfo {author} {\bibfnamefont {S.}~\bibnamefont {Gaucher}}, \bibinfo
  {author} {\bibfnamefont {G.}~\bibnamefont {Gervais}}, \bibinfo {author}
  {\bibfnamefont {L.~N.}\ \bibnamefont {Pfeiffer}}, \ and\ \bibinfo {author}
  {\bibfnamefont {K.~W.}\ \bibnamefont {West}},\ }\bibfield  {title} {\enquote
  {\bibinfo {title} {Specific heat and entropy of fractional quantum hall
  states in the second landau level},}\ }\href {\doibase
  10.1103/PhysRevB.95.201306} {\bibfield  {journal} {\bibinfo  {journal} {Phys.
  Rev. B}\ }\textbf {\bibinfo {volume} {95}},\ \bibinfo {pages} {201306}
  (\bibinfo {year} {2017})}\BibitemShut {NoStop}%
\bibitem [{\citenamefont {Tarucha}\ \emph {et~al.}(1996)\citenamefont
  {Tarucha}, \citenamefont {Austing}, \citenamefont {Honda}, \citenamefont
  {van~der Hage},\ and\ \citenamefont {Kouwenhoven}}]{Tarucha1996}%
  \BibitemOpen
  \bibfield  {author} {\bibinfo {author} {\bibfnamefont {S.}~\bibnamefont
  {Tarucha}}, \bibinfo {author} {\bibfnamefont {D.~G.}\ \bibnamefont
  {Austing}}, \bibinfo {author} {\bibfnamefont {T.}~\bibnamefont {Honda}},
  \bibinfo {author} {\bibfnamefont {R.~J.}\ \bibnamefont {van~der Hage}}, \
  and\ \bibinfo {author} {\bibfnamefont {L.~P.}\ \bibnamefont {Kouwenhoven}},\
  }\bibfield  {title} {\enquote {\bibinfo {title} {{Shell filling and spin
  effects in a few electron quantum dot}},}\ }\href {\doibase
  10.1103/PhysRevLett.77.3613} {\bibfield  {journal} {\bibinfo  {journal}
  {Phys. Rev. Lett.}\ }\textbf {\bibinfo {volume} {77}},\ \bibinfo {pages}
  {3613--3616} (\bibinfo {year} {1996})}\BibitemShut {NoStop}%
\bibitem [{\citenamefont {Ciorga}\ \emph {et~al.}(2000)\citenamefont {Ciorga},
  \citenamefont {Sachrajda}, \citenamefont {Hawrylak}, \citenamefont {Gould},
  \citenamefont {Zawadzki}, \citenamefont {Jullian}, \citenamefont {Feng},\
  and\ \citenamefont {Wasilewski}}]{Ciorga2000}%
  \BibitemOpen
  \bibfield  {author} {\bibinfo {author} {\bibfnamefont {M.}~\bibnamefont
  {Ciorga}}, \bibinfo {author} {\bibfnamefont {A.}~\bibnamefont {Sachrajda}},
  \bibinfo {author} {\bibfnamefont {P.}~\bibnamefont {Hawrylak}}, \bibinfo
  {author} {\bibfnamefont {C.}~\bibnamefont {Gould}}, \bibinfo {author}
  {\bibfnamefont {P.}~\bibnamefont {Zawadzki}}, \bibinfo {author}
  {\bibfnamefont {S.}~\bibnamefont {Jullian}}, \bibinfo {author} {\bibfnamefont
  {Y.}~\bibnamefont {Feng}}, \ and\ \bibinfo {author} {\bibfnamefont
  {Z.}~\bibnamefont {Wasilewski}},\ }\bibfield  {title} {\enquote {\bibinfo
  {title} {{Addition spectrum of a lateral dot from Coulomb and spin-blockade
  spectroscopy}},}\ }\href {\doibase 10.1103/PhysRevB.61.R16315} {\bibfield
  {journal} {\bibinfo  {journal} {Phys. Rev. B}\ }\textbf {\bibinfo {volume}
  {61}},\ \bibinfo {pages} {R16315--R16318} (\bibinfo {year}
  {2000})}\BibitemShut {NoStop}%
\bibitem [{\citenamefont {Duncan}\ \emph {et~al.}(2000)\citenamefont {Duncan},
  \citenamefont {Goldhaber-Gordon}, \citenamefont {Westervelt}, \citenamefont
  {Maranowski},\ and\ \citenamefont {Gossard}}]{Duncan2000}%
  \BibitemOpen
  \bibfield  {author} {\bibinfo {author} {\bibfnamefont {D.~S.}\ \bibnamefont
  {Duncan}}, \bibinfo {author} {\bibfnamefont {D.}~\bibnamefont
  {Goldhaber-Gordon}}, \bibinfo {author} {\bibfnamefont {R.~M.}\ \bibnamefont
  {Westervelt}}, \bibinfo {author} {\bibfnamefont {K.~D.}\ \bibnamefont
  {Maranowski}}, \ and\ \bibinfo {author} {\bibfnamefont {A.~C.}\ \bibnamefont
  {Gossard}},\ }\bibfield  {title} {\enquote {\bibinfo {title}
  {{Coulomb-blockade spectroscopy on a small quantum dot in a parallel magnetic
  field}},}\ }\href {\doibase 10.1063/1.1313812} {\bibfield  {journal}
  {\bibinfo  {journal} {Appl. Phys. Lett.}\ }\textbf {\bibinfo {volume} {77}},\
  \bibinfo {pages} {2183--2185} (\bibinfo {year} {2000})}\BibitemShut {NoStop}%
\bibitem [{\citenamefont {Lindemann}\ \emph {et~al.}(2002)\citenamefont
  {Lindemann}, \citenamefont {Ihn}, \citenamefont {Heinzel}, \citenamefont
  {Zwerger}, \citenamefont {Ensslin}, \citenamefont {Maranowski},\ and\
  \citenamefont {Gossard}}]{Lindemann2002}%
  \BibitemOpen
  \bibfield  {author} {\bibinfo {author} {\bibfnamefont {S.}~\bibnamefont
  {Lindemann}}, \bibinfo {author} {\bibfnamefont {T.}~\bibnamefont {Ihn}},
  \bibinfo {author} {\bibfnamefont {T.}~\bibnamefont {Heinzel}}, \bibinfo
  {author} {\bibfnamefont {W.}~\bibnamefont {Zwerger}}, \bibinfo {author}
  {\bibfnamefont {K.}~\bibnamefont {Ensslin}}, \bibinfo {author} {\bibfnamefont
  {K.}~\bibnamefont {Maranowski}}, \ and\ \bibinfo {author} {\bibfnamefont
  {A.~C.}\ \bibnamefont {Gossard}},\ }\bibfield  {title} {\enquote {\bibinfo
  {title} {Stability of spin states in quantum dots},}\ }\href {\doibase
  10.1103/PhysRevB.66.195314} {\bibfield  {journal} {\bibinfo  {journal} {Phys.
  Rev. B}\ }\textbf {\bibinfo {volume} {66}},\ \bibinfo {pages} {195314}
  (\bibinfo {year} {2002})}\BibitemShut {NoStop}%
\bibitem [{\citenamefont {Potok}\ \emph {et~al.}(2003)\citenamefont {Potok},
  \citenamefont {Folk}, \citenamefont {Marcus}, \citenamefont {Umansky},
  \citenamefont {Hanson},\ and\ \citenamefont {Gossard}}]{Potok2003}%
  \BibitemOpen
  \bibfield  {author} {\bibinfo {author} {\bibfnamefont {R.~M.}\ \bibnamefont
  {Potok}}, \bibinfo {author} {\bibfnamefont {J.~A.}\ \bibnamefont {Folk}},
  \bibinfo {author} {\bibfnamefont {C.~M.}\ \bibnamefont {Marcus}}, \bibinfo
  {author} {\bibfnamefont {V.}~\bibnamefont {Umansky}}, \bibinfo {author}
  {\bibfnamefont {M.}~\bibnamefont {Hanson}}, \ and\ \bibinfo {author}
  {\bibfnamefont {A.~C.}\ \bibnamefont {Gossard}},\ }\bibfield  {title}
  {\enquote {\bibinfo {title} {Spin and polarized current from coulomb
  blockaded quantum dots},}\ }\href {\doibase 10.1103/PhysRevLett.91.016802}
  {\bibfield  {journal} {\bibinfo  {journal} {Phys. Rev. Lett.}\ }\textbf
  {\bibinfo {volume} {91}},\ \bibinfo {pages} {016802} (\bibinfo {year}
  {2003})}\BibitemShut {NoStop}%
\bibitem [{\citenamefont {Hofmann}\ \emph {et~al.}(2016)\citenamefont
  {Hofmann}, \citenamefont {Maisi}, \citenamefont {Gold}, \citenamefont
  {Kr\"ahenmann}, \citenamefont {R\"ossler}, \citenamefont {Basset},
  \citenamefont {M\"arki}, \citenamefont {Reichl}, \citenamefont {Wegscheider},
  \citenamefont {Ensslin},\ and\ \citenamefont {Ihn}}]{Hofmann2016}%
  \BibitemOpen
  \bibfield  {author} {\bibinfo {author} {\bibfnamefont {A.}~\bibnamefont
  {Hofmann}}, \bibinfo {author} {\bibfnamefont {V.~F.}\ \bibnamefont {Maisi}},
  \bibinfo {author} {\bibfnamefont {C.}~\bibnamefont {Gold}}, \bibinfo {author}
  {\bibfnamefont {T.}~\bibnamefont {Kr\"ahenmann}}, \bibinfo {author}
  {\bibfnamefont {C.}~\bibnamefont {R\"ossler}}, \bibinfo {author}
  {\bibfnamefont {J.}~\bibnamefont {Basset}}, \bibinfo {author} {\bibfnamefont
  {P.}~\bibnamefont {M\"arki}}, \bibinfo {author} {\bibfnamefont
  {C.}~\bibnamefont {Reichl}}, \bibinfo {author} {\bibfnamefont
  {W.}~\bibnamefont {Wegscheider}}, \bibinfo {author} {\bibfnamefont
  {K.}~\bibnamefont {Ensslin}}, \ and\ \bibinfo {author} {\bibfnamefont
  {T.}~\bibnamefont {Ihn}},\ }\bibfield  {title} {\enquote {\bibinfo {title}
  {Measuring the degeneracy of discrete energy levels using a
  $\mathrm{GaAs}/\mathrm{AlGaAs}$ quantum dot},}\ }\href {\doibase
  10.1103/PhysRevLett.117.206803} {\bibfield  {journal} {\bibinfo  {journal}
  {Phys. Rev. Lett.}\ }\textbf {\bibinfo {volume} {117}},\ \bibinfo {pages}
  {206803} (\bibinfo {year} {2016})}\BibitemShut {NoStop}%
\bibitem [{\citenamefont {Cooper}\ and\ \citenamefont
  {Stern}(2009)}]{Cooper2009}%
  \BibitemOpen
  \bibfield  {author} {\bibinfo {author} {\bibfnamefont {N.~R.}\ \bibnamefont
  {Cooper}}\ and\ \bibinfo {author} {\bibfnamefont {Ady}\ \bibnamefont
  {Stern}},\ }\bibfield  {title} {\enquote {\bibinfo {title} {Observable bulk
  signatures of non-abelian quantum hall states},}\ }\href {\doibase
  10.1103/PhysRevLett.102.176807} {\bibfield  {journal} {\bibinfo  {journal}
  {Phys. Rev. Lett.}\ }\textbf {\bibinfo {volume} {102}},\ \bibinfo {pages}
  {176807} (\bibinfo {year} {2009})}\BibitemShut {NoStop}%
\bibitem [{\citenamefont {Ben-Shach}\ \emph {et~al.}(2013)\citenamefont
  {Ben-Shach}, \citenamefont {Laumann}, \citenamefont {Neder}, \citenamefont
  {Yacoby},\ and\ \citenamefont {Halperin}}]{Ben-Shach2013}%
  \BibitemOpen
  \bibfield  {author} {\bibinfo {author} {\bibfnamefont {G.}~\bibnamefont
  {Ben-Shach}}, \bibinfo {author} {\bibfnamefont {C.~R.}\ \bibnamefont
  {Laumann}}, \bibinfo {author} {\bibfnamefont {I.}~\bibnamefont {Neder}},
  \bibinfo {author} {\bibfnamefont {A.}~\bibnamefont {Yacoby}}, \ and\ \bibinfo
  {author} {\bibfnamefont {B.~I.}\ \bibnamefont {Halperin}},\ }\bibfield
  {title} {\enquote {\bibinfo {title} {Detecting non-abelian anyons by charging
  spectroscopy},}\ }\href {\doibase 10.1103/PhysRevLett.110.106805} {\bibfield
  {journal} {\bibinfo  {journal} {Phys. Rev. Lett.}\ }\textbf {\bibinfo
  {volume} {110}},\ \bibinfo {pages} {106805} (\bibinfo {year}
  {2013})}\BibitemShut {NoStop}%
\bibitem [{\citenamefont {Smirnov}(2015)}]{Smirnov2015}%
  \BibitemOpen
  \bibfield  {author} {\bibinfo {author} {\bibfnamefont {Sergey}\ \bibnamefont
  {Smirnov}},\ }\bibfield  {title} {\enquote {\bibinfo {title} {Majorana
  tunneling entropy},}\ }\href {\doibase 10.1103/PhysRevB.92.195312} {\bibfield
   {journal} {\bibinfo  {journal} {Phys. Rev. B}\ }\textbf {\bibinfo {volume}
  {92}},\ \bibinfo {pages} {195312} (\bibinfo {year} {2015})}\BibitemShut
  {NoStop}%
\bibitem [{\citenamefont {Hou}\ \emph {et~al.}(2012)\citenamefont {Hou},
  \citenamefont {Shtengel}, \citenamefont {Refael},\ and\ \citenamefont
  {Goldbart}}]{Hou2012}%
  \BibitemOpen
  \bibfield  {author} {\bibinfo {author} {\bibfnamefont {C-Y}\ \bibnamefont
  {Hou}}, \bibinfo {author} {\bibfnamefont {K}~\bibnamefont {Shtengel}},
  \bibinfo {author} {\bibfnamefont {G}~\bibnamefont {Refael}}, \ and\ \bibinfo
  {author} {\bibfnamefont {P~M}\ \bibnamefont {Goldbart}},\ }\bibfield  {title}
  {\enquote {\bibinfo {title} {Ettingshausen effect due to majorana modes},}\
  }\href {http://stacks.iop.org/1367-2630/14/i=10/a=105005} {\bibfield
  {journal} {\bibinfo  {journal} {New Journal of Physics}\ }\textbf {\bibinfo
  {volume} {14}},\ \bibinfo {pages} {105005} (\bibinfo {year}
  {2012})}\BibitemShut {NoStop}%
\bibitem [{\citenamefont {Alkurtass}\ \emph {et~al.}(2016)\citenamefont
  {Alkurtass}, \citenamefont {Bayat}, \citenamefont {Affleck}, \citenamefont
  {Bose}, \citenamefont {Johannesson}, \citenamefont {Sodano}, \citenamefont
  {S\o{}rensen},\ and\ \citenamefont {Le~Hur}}]{Alkurtass2016}%
  \BibitemOpen
  \bibfield  {author} {\bibinfo {author} {\bibfnamefont {Bedoor}\ \bibnamefont
  {Alkurtass}}, \bibinfo {author} {\bibfnamefont {Abolfazl}\ \bibnamefont
  {Bayat}}, \bibinfo {author} {\bibfnamefont {Ian}\ \bibnamefont {Affleck}},
  \bibinfo {author} {\bibfnamefont {Sougato}\ \bibnamefont {Bose}}, \bibinfo
  {author} {\bibfnamefont {Henrik}\ \bibnamefont {Johannesson}}, \bibinfo
  {author} {\bibfnamefont {Pasquale}\ \bibnamefont {Sodano}}, \bibinfo {author}
  {\bibfnamefont {Erik~S.}\ \bibnamefont {S\o{}rensen}}, \ and\ \bibinfo
  {author} {\bibfnamefont {Karyn}\ \bibnamefont {Le~Hur}},\ }\bibfield  {title}
  {\enquote {\bibinfo {title} {Entanglement structure of the two-channel kondo
  model},}\ }\href {\doibase 10.1103/PhysRevB.93.081106} {\bibfield  {journal}
  {\bibinfo  {journal} {Phys. Rev. B}\ }\textbf {\bibinfo {volume} {93}},\
  \bibinfo {pages} {081106} (\bibinfo {year} {2016})}\BibitemShut {NoStop}%
\bibitem [{\citenamefont {Elzerman}\ \emph {et~al.}(2004)\citenamefont
  {Elzerman}, \citenamefont {Hanson}, \citenamefont {van Beveren},
  \citenamefont {Witkamp}, \citenamefont {Kouwenhoven},\ and\ \citenamefont
  {Vandersypen}}]{Elzerman2004}%
  \BibitemOpen
  \bibfield  {author} {\bibinfo {author} {\bibfnamefont {J.~M.}\ \bibnamefont
  {Elzerman}}, \bibinfo {author} {\bibfnamefont {R.}~\bibnamefont {Hanson}},
  \bibinfo {author} {\bibfnamefont {L.~H.~Willems}\ \bibnamefont {van
  Beveren}}, \bibinfo {author} {\bibfnamefont {B.}~\bibnamefont {Witkamp}},
  \bibinfo {author} {\bibfnamefont {L.P.}\ \bibnamefont {Kouwenhoven}}, \ and\
  \bibinfo {author} {\bibfnamefont {L.~M.~K.}\ \bibnamefont {Vandersypen}},\
  }\bibfield  {title} {\enquote {\bibinfo {title} {{Single-shot read-out of an
  individual electron spin in a quantum dot}},}\ }\href {\doibase
  10.1038/nature02693} {\bibfield  {journal} {\bibinfo  {journal} {Nature}\
  }\textbf {\bibinfo {volume} {430}},\ \bibinfo {pages} {431--435} (\bibinfo
  {year} {2004})}\BibitemShut {NoStop}%
\bibitem [{\citenamefont {Ono}\ and\ \citenamefont {Tarucha}(2004)}]{Ono2004}%
  \BibitemOpen
  \bibfield  {author} {\bibinfo {author} {\bibfnamefont {Keiji}\ \bibnamefont
  {Ono}}\ and\ \bibinfo {author} {\bibfnamefont {Seigo}\ \bibnamefont
  {Tarucha}},\ }\bibfield  {title} {\enquote {\bibinfo {title}
  {Nuclear-spin-induced oscillatory current in spin-blockaded quantum dots},}\
  }\href {\doibase 10.1103/PhysRevLett.92.256803} {\bibfield  {journal}
  {\bibinfo  {journal} {Phys. Rev. Lett.}\ }\textbf {\bibinfo {volume} {92}},\
  \bibinfo {pages} {256803} (\bibinfo {year} {2004})}\BibitemShut {NoStop}%
\bibitem [{\citenamefont {Venkatachalam}\ \emph {et~al.}(2011)\citenamefont
  {Venkatachalam}, \citenamefont {Yacoby}, \citenamefont {Pfeiffer},\ and\
  \citenamefont {West}}]{Venkatachalam2011}%
  \BibitemOpen
  \bibfield  {author} {\bibinfo {author} {\bibfnamefont {Vivek}\ \bibnamefont
  {Venkatachalam}}, \bibinfo {author} {\bibfnamefont {Amir}\ \bibnamefont
  {Yacoby}}, \bibinfo {author} {\bibfnamefont {Loren}\ \bibnamefont
  {Pfeiffer}}, \ and\ \bibinfo {author} {\bibfnamefont {Ken}\ \bibnamefont
  {West}},\ }\bibfield  {title} {\enquote {\bibinfo {title} {{Local charge of
  the $\nu$ = 5/2 fractional quantum Hall state}},}\ }\href {\doibase
  10.1038/nature09680} {\bibfield  {journal} {\bibinfo  {journal} {Nature}\
  }\textbf {\bibinfo {volume} {469}},\ \bibinfo {pages} {185--188} (\bibinfo
  {year} {2011})}\BibitemShut {NoStop}%
\bibitem [{\citenamefont {Field}\ \emph {et~al.}(1993)\citenamefont {Field},
  \citenamefont {Smith}, \citenamefont {Pepper}, \citenamefont {Ritchie},
  \citenamefont {Frost}, \citenamefont {Jones},\ and\ \citenamefont
  {Hasko}}]{Field1993}%
  \BibitemOpen
  \bibfield  {author} {\bibinfo {author} {\bibfnamefont {M.}~\bibnamefont
  {Field}}, \bibinfo {author} {\bibfnamefont {C.~G.}\ \bibnamefont {Smith}},
  \bibinfo {author} {\bibfnamefont {M.}~\bibnamefont {Pepper}}, \bibinfo
  {author} {\bibfnamefont {D.~A.}\ \bibnamefont {Ritchie}}, \bibinfo {author}
  {\bibfnamefont {J.~E.~F.}\ \bibnamefont {Frost}}, \bibinfo {author}
  {\bibfnamefont {G.~A.~C.}\ \bibnamefont {Jones}}, \ and\ \bibinfo {author}
  {\bibfnamefont {D.~G.}\ \bibnamefont {Hasko}},\ }\bibfield  {title} {\enquote
  {\bibinfo {title} {Measurements of coulomb blockade with a noninvasive
  voltage probe},}\ }\href {\doibase 10.1103/PhysRevLett.70.1311} {\bibfield
  {journal} {\bibinfo  {journal} {Phys. Rev. Lett.}\ }\textbf {\bibinfo
  {volume} {70}},\ \bibinfo {pages} {1311--1314} (\bibinfo {year}
  {1993})}\BibitemShut {NoStop}%
\bibitem [{\citenamefont {Staring}\ \emph {et~al.}(1993)\citenamefont
  {Staring}, \citenamefont {Molenkamp}, \citenamefont {Alphenaar},
  \citenamefont {van Houten}, \citenamefont {Buyk}, \citenamefont {Mabesoone},
  \citenamefont {Beenakker},\ and\ \citenamefont {Foxon}}]{Staring2007}%
  \BibitemOpen
  \bibfield  {author} {\bibinfo {author} {\bibfnamefont {A.~A.~M.}\
  \bibnamefont {Staring}}, \bibinfo {author} {\bibfnamefont {L.~W.}\
  \bibnamefont {Molenkamp}}, \bibinfo {author} {\bibfnamefont {B.~W.}\
  \bibnamefont {Alphenaar}}, \bibinfo {author} {\bibfnamefont {H.}~\bibnamefont
  {van Houten}}, \bibinfo {author} {\bibfnamefont {O.~J.~A.}\ \bibnamefont
  {Buyk}}, \bibinfo {author} {\bibfnamefont {M.~A.~A.}\ \bibnamefont
  {Mabesoone}}, \bibinfo {author} {\bibfnamefont {C.~W.~J.}\ \bibnamefont
  {Beenakker}}, \ and\ \bibinfo {author} {\bibfnamefont {C.~T.}\ \bibnamefont
  {Foxon}},\ }\bibfield  {title} {\enquote {\bibinfo {title} {Coulomb-blockade
  oscillations in the thermopower of a quantum dot},}\ }\href
  {http://stacks.iop.org/0295-5075/22/i=1/a=011} {\bibfield  {journal}
  {\bibinfo  {journal} {EPL (Europhysics Letters)}\ }\textbf {\bibinfo {volume}
  {22}},\ \bibinfo {pages} {57} (\bibinfo {year} {1993})}\BibitemShut {NoStop}%
\bibitem [{\citenamefont {Thierschmann}\ \emph {et~al.}(2015)\citenamefont
  {Thierschmann}, \citenamefont {S{\'{a}}nchez}, \citenamefont {Sothmann},
  \citenamefont {Arnold}, \citenamefont {Heyn}, \citenamefont {Hansen},
  \citenamefont {Buhmann},\ and\ \citenamefont {Molenkamp}}]{Thierschmann2015}%
  \BibitemOpen
  \bibfield  {author} {\bibinfo {author} {\bibfnamefont {Holger}\ \bibnamefont
  {Thierschmann}}, \bibinfo {author} {\bibfnamefont {Rafael}\ \bibnamefont
  {S{\'{a}}nchez}}, \bibinfo {author} {\bibfnamefont {Bj{\"{o}}rn}\
  \bibnamefont {Sothmann}}, \bibinfo {author} {\bibfnamefont {Fabian}\
  \bibnamefont {Arnold}}, \bibinfo {author} {\bibfnamefont {Christian}\
  \bibnamefont {Heyn}}, \bibinfo {author} {\bibfnamefont {Wolfgang}\
  \bibnamefont {Hansen}}, \bibinfo {author} {\bibfnamefont {Hartmut}\
  \bibnamefont {Buhmann}}, \ and\ \bibinfo {author} {\bibfnamefont
  {Laurens~W.}\ \bibnamefont {Molenkamp}},\ }\bibfield  {title} {\enquote
  {\bibinfo {title} {{Three-terminal energy harvester with coupled quantum
  dots}},}\ }\href {\doibase 10.1038/nnano.2015.176} {\bibfield  {journal}
  {\bibinfo  {journal} {Nature Nanotechnology}\ }\textbf {\bibinfo {volume}
  {10}},\ \bibinfo {pages} {854--858} (\bibinfo {year} {2015})}\BibitemShut
  {NoStop}%
\bibitem [{\citenamefont {Beenakker}(1991)}]{Beenakker1991}%
  \BibitemOpen
  \bibfield  {author} {\bibinfo {author} {\bibfnamefont {C.~W.~J.}\
  \bibnamefont {Beenakker}},\ }\bibfield  {title} {\enquote {\bibinfo {title}
  {Theory of coulomb-blockade oscillations in the conductance of a quantum
  dot},}\ }\href {\doibase 10.1103/PhysRevB.44.1646} {\bibfield  {journal}
  {\bibinfo  {journal} {Phys. Rev. B}\ }\textbf {\bibinfo {volume} {44}},\
  \bibinfo {pages} {1646--1656} (\bibinfo {year} {1991})}\BibitemShut {NoStop}%
\bibitem [{\citenamefont {Gustavsson}\ \emph {et~al.}(2009)\citenamefont
  {Gustavsson}, \citenamefont {Leturcq}, \citenamefont {Studer}, \citenamefont
  {Shorubalko}, \citenamefont {Ihn}, \citenamefont {Ensslin}, \citenamefont
  {Driscoll},\ and\ \citenamefont {Gossard}}]{Gustavsson2009}%
  \BibitemOpen
  \bibfield  {author} {\bibinfo {author} {\bibfnamefont {S.}~\bibnamefont
  {Gustavsson}}, \bibinfo {author} {\bibfnamefont {R.}~\bibnamefont {Leturcq}},
  \bibinfo {author} {\bibfnamefont {M.}~\bibnamefont {Studer}}, \bibinfo
  {author} {\bibfnamefont {I.}~\bibnamefont {Shorubalko}}, \bibinfo {author}
  {\bibfnamefont {T.}~\bibnamefont {Ihn}}, \bibinfo {author} {\bibfnamefont
  {K.}~\bibnamefont {Ensslin}}, \bibinfo {author} {\bibfnamefont {D.~C.}\
  \bibnamefont {Driscoll}}, \ and\ \bibinfo {author} {\bibfnamefont {A.~C.}\
  \bibnamefont {Gossard}},\ }\bibfield  {title} {\enquote {\bibinfo {title}
  {{Electron counting in quantum dots}},}\ }\href {\doibase
  10.1016/j.surfrep.2009.02.001} {\bibfield  {journal} {\bibinfo  {journal}
  {Surface Science Reports}\ }\textbf {\bibinfo {volume} {64}},\ \bibinfo
  {pages} {191--232} (\bibinfo {year} {2009})}\BibitemShut {NoStop}%
\bibitem [{\citenamefont {Cockins}\ \emph {et~al.}(2010)\citenamefont
  {Cockins}, \citenamefont {Miyahara}, \citenamefont {Bennett}, \citenamefont
  {Clerk}, \citenamefont {Studenikin}, \citenamefont {Poole}, \citenamefont
  {Sachrajda},\ and\ \citenamefont {Grutter}}]{Cockins2010}%
  \BibitemOpen
  \bibfield  {author} {\bibinfo {author} {\bibfnamefont {Lynda}\ \bibnamefont
  {Cockins}}, \bibinfo {author} {\bibfnamefont {Yoichi}\ \bibnamefont
  {Miyahara}}, \bibinfo {author} {\bibfnamefont {Steven~D.}\ \bibnamefont
  {Bennett}}, \bibinfo {author} {\bibfnamefont {Aashish~A.}\ \bibnamefont
  {Clerk}}, \bibinfo {author} {\bibfnamefont {Sergei}\ \bibnamefont
  {Studenikin}}, \bibinfo {author} {\bibfnamefont {Philip}\ \bibnamefont
  {Poole}}, \bibinfo {author} {\bibfnamefont {Andrew}\ \bibnamefont
  {Sachrajda}}, \ and\ \bibinfo {author} {\bibfnamefont {Peter}\ \bibnamefont
  {Grutter}},\ }\bibfield  {title} {\enquote {\bibinfo {title} {Energy levels
  of few-electron quantum dots imaged and characterized by atomic force
  microscopy},}\ }\href {\doibase 10.1073/pnas.0912716107} {\bibfield
  {journal} {\bibinfo  {journal} {Proceedings of the National Academy of
  Sciences}\ }\textbf {\bibinfo {volume} {107}},\ \bibinfo {pages} {9496--9501}
  (\bibinfo {year} {2010})}\BibitemShut {NoStop}%
\bibitem [{\citenamefont {Bennett}\ \emph {et~al.}(2010)\citenamefont
  {Bennett}, \citenamefont {Cockins}, \citenamefont {Miyahara}, \citenamefont
  {Gr\"utter},\ and\ \citenamefont {Clerk}}]{Bennett2010}%
  \BibitemOpen
  \bibfield  {author} {\bibinfo {author} {\bibfnamefont {Steven~D.}\
  \bibnamefont {Bennett}}, \bibinfo {author} {\bibfnamefont {Lynda}\
  \bibnamefont {Cockins}}, \bibinfo {author} {\bibfnamefont {Yoichi}\
  \bibnamefont {Miyahara}}, \bibinfo {author} {\bibfnamefont {Peter}\
  \bibnamefont {Gr\"utter}}, \ and\ \bibinfo {author} {\bibfnamefont
  {Aashish~A.}\ \bibnamefont {Clerk}},\ }\bibfield  {title} {\enquote {\bibinfo
  {title} {Strong electromechanical coupling of an atomic force microscope
  cantilever to a quantum dot},}\ }\href {\doibase
  10.1103/PhysRevLett.104.017203} {\bibfield  {journal} {\bibinfo  {journal}
  {Phys. Rev. Lett.}\ }\textbf {\bibinfo {volume} {104}},\ \bibinfo {pages}
  {017203} (\bibinfo {year} {2010})}\BibitemShut {NoStop}%
\bibitem [{\citenamefont {Beckel}\ \emph {et~al.}(2014)\citenamefont {Beckel},
  \citenamefont {Kurzmann}, \citenamefont {Geller}, \citenamefont {Ludwig},
  \citenamefont {Wieck}, \citenamefont {König},\ and\ \citenamefont
  {Lorke}}]{Beckel2014}%
  \BibitemOpen
  \bibfield  {author} {\bibinfo {author} {\bibfnamefont {A.}~\bibnamefont
  {Beckel}}, \bibinfo {author} {\bibfnamefont {A.}~\bibnamefont {Kurzmann}},
  \bibinfo {author} {\bibfnamefont {M.}~\bibnamefont {Geller}}, \bibinfo
  {author} {\bibfnamefont {A.}~\bibnamefont {Ludwig}}, \bibinfo {author}
  {\bibfnamefont {A.~D.}\ \bibnamefont {Wieck}}, \bibinfo {author}
  {\bibfnamefont {J.}~\bibnamefont {König}}, \ and\ \bibinfo {author}
  {\bibfnamefont {A.}~\bibnamefont {Lorke}},\ }\bibfield  {title} {\enquote
  {\bibinfo {title} {Asymmetry of charge relaxation times in quantum dots: The
  influence of degeneracy},}\ }\href
  {http://stacks.iop.org/0295-5075/106/i=4/a=47002} {\bibfield  {journal}
  {\bibinfo  {journal} {EPL (Europhysics Letters)}\ }\textbf {\bibinfo {volume}
  {106}},\ \bibinfo {pages} {47002} (\bibinfo {year} {2014})}\BibitemShut
  {NoStop}%
\bibitem [{\citenamefont {Mittal}\ \emph {et~al.}(1996)\citenamefont {Mittal},
  \citenamefont {Wheeler}, \citenamefont {Keller}, \citenamefont {Prober},\
  and\ \citenamefont {Sacks}}]{Mittal1996}%
  \BibitemOpen
  \bibfield  {author} {\bibinfo {author} {\bibfnamefont {A.}~\bibnamefont
  {Mittal}}, \bibinfo {author} {\bibfnamefont {R.~G.}\ \bibnamefont {Wheeler}},
  \bibinfo {author} {\bibfnamefont {M.~W.}\ \bibnamefont {Keller}}, \bibinfo
  {author} {\bibfnamefont {D.~E.}\ \bibnamefont {Prober}}, \ and\ \bibinfo
  {author} {\bibfnamefont {R.~N.}\ \bibnamefont {Sacks}},\ }\bibfield  {title}
  {\enquote {\bibinfo {title} {{Electron-phonon scattering rates in GaAs/AlGaAs
  2DEG samples below 0.5 K}},}\ }\href {\doibase 10.1016/0039-6028(96)00464-5}
  {\bibfield  {journal} {\bibinfo  {journal} {Surface Science}\ }\textbf
  {\bibinfo {volume} {361-362}},\ \bibinfo {pages} {537--541} (\bibinfo {year}
  {1996})}\BibitemShut {NoStop}%
\bibitem [{\citenamefont {Cronenwett}\ \emph {et~al.}(1998)\citenamefont
  {Cronenwett}, \citenamefont {Oosterkamp},\ and\ \citenamefont
  {Kouwenhoven}}]{Cronenwett1998}%
  \BibitemOpen
  \bibfield  {author} {\bibinfo {author} {\bibfnamefont {Sara~M.}\ \bibnamefont
  {Cronenwett}}, \bibinfo {author} {\bibfnamefont {Tjerk~H.}\ \bibnamefont
  {Oosterkamp}}, \ and\ \bibinfo {author} {\bibfnamefont {Leo~P.}\ \bibnamefont
  {Kouwenhoven}},\ }\bibfield  {title} {\enquote {\bibinfo {title} {A tunable
  kondo effect in quantum dots},}\ }\href {\doibase
  10.1126/science.281.5376.540} {\bibfield  {journal} {\bibinfo  {journal}
  {Science}\ }\textbf {\bibinfo {volume} {281}},\ \bibinfo {pages} {540--544}
  (\bibinfo {year} {1998})}\BibitemShut {NoStop}%
\bibitem [{\citenamefont {Hanson}\ \emph {et~al.}(2003)\citenamefont {Hanson},
  \citenamefont {Witkamp}, \citenamefont {Vandersypen}, \citenamefont {van
  Beveren}, \citenamefont {Elzerman},\ and\ \citenamefont
  {Kouwenhoven}}]{Hanson2003}%
  \BibitemOpen
  \bibfield  {author} {\bibinfo {author} {\bibfnamefont {R.}~\bibnamefont
  {Hanson}}, \bibinfo {author} {\bibfnamefont {B.}~\bibnamefont {Witkamp}},
  \bibinfo {author} {\bibfnamefont {L.~M.~K.}\ \bibnamefont {Vandersypen}},
  \bibinfo {author} {\bibfnamefont {L.~H.~Willems}\ \bibnamefont {van
  Beveren}}, \bibinfo {author} {\bibfnamefont {J.~M.}\ \bibnamefont
  {Elzerman}}, \ and\ \bibinfo {author} {\bibfnamefont {L.~P.}\ \bibnamefont
  {Kouwenhoven}},\ }\bibfield  {title} {\enquote {\bibinfo {title} {Zeeman
  energy and spin relaxation in a one-electron quantum dot},}\ }\href {\doibase
  10.1103/PhysRevLett.91.196802} {\bibfield  {journal} {\bibinfo  {journal}
  {Phys. Rev. Lett.}\ }\textbf {\bibinfo {volume} {91}},\ \bibinfo {pages}
  {196802} (\bibinfo {year} {2003})}\BibitemShut {NoStop}%
\bibitem [{\citenamefont {Zumb\"uhl}\ \emph {et~al.}(2004)\citenamefont
  {Zumb\"uhl}, \citenamefont {Marcus}, \citenamefont {Hanson},\ and\
  \citenamefont {Gossard}}]{Zumbuhl2004}%
  \BibitemOpen
  \bibfield  {author} {\bibinfo {author} {\bibfnamefont {D.~M.}\ \bibnamefont
  {Zumb\"uhl}}, \bibinfo {author} {\bibfnamefont {C.~M.}\ \bibnamefont
  {Marcus}}, \bibinfo {author} {\bibfnamefont {M.~P.}\ \bibnamefont {Hanson}},
  \ and\ \bibinfo {author} {\bibfnamefont {A.~C.}\ \bibnamefont {Gossard}},\
  }\bibfield  {title} {\enquote {\bibinfo {title} {Cotunneling spectroscopy in
  few-electron quantum dots},}\ }\href {\doibase 10.1103/PhysRevLett.93.256801}
  {\bibfield  {journal} {\bibinfo  {journal} {Phys. Rev. Lett.}\ }\textbf
  {\bibinfo {volume} {93}},\ \bibinfo {pages} {256801} (\bibinfo {year}
  {2004})}\BibitemShut {NoStop}%
\bibitem [{\citenamefont {Szafran}\ \emph {et~al.}(2004)\citenamefont
  {Szafran}, \citenamefont {Peeters}, \citenamefont {Bednarek},\ and\
  \citenamefont {Adamowski}}]{Szafran2004}%
  \BibitemOpen
  \bibfield  {author} {\bibinfo {author} {\bibfnamefont {B.}~\bibnamefont
  {Szafran}}, \bibinfo {author} {\bibfnamefont {F.~M.}\ \bibnamefont
  {Peeters}}, \bibinfo {author} {\bibfnamefont {S.}~\bibnamefont {Bednarek}}, \
  and\ \bibinfo {author} {\bibfnamefont {J.}~\bibnamefont {Adamowski}},\
  }\bibfield  {title} {\enquote {\bibinfo {title} {In-plane
  magnetic-field-induced wigner crystallization in a two-electron quantum
  dot},}\ }\href {\doibase 10.1103/PhysRevB.70.235335} {\bibfield  {journal}
  {\bibinfo  {journal} {Phys. Rev. B}\ }\textbf {\bibinfo {volume} {70}},\
  \bibinfo {pages} {235335} (\bibinfo {year} {2004})}\BibitemShut {NoStop}%
\bibitem [{\citenamefont {Landau}\ and\ \citenamefont
  {Lifshitz}(1980)}]{Landau1980}%
  \BibitemOpen
  \bibfield  {author} {\bibinfo {author} {\bibfnamefont {L.D.}\ \bibnamefont
  {Landau}}\ and\ \bibinfo {author} {\bibfnamefont {E.M.}\ \bibnamefont
  {Lifshitz}},\ }\bibfield  {title} {\enquote {\bibinfo {title} {Chapter v -
  the fermi and bose distributions},}\ }in\ \href {\doibase
  https://doi.org/10.1016/B978-0-08-057046-4.50012-9} {\emph {\bibinfo
  {booktitle} {Statistical Physics (Third Edition)}}}\ (\bibinfo  {publisher}
  {Butterworth-Heinemann},\ \bibinfo {address} {Oxford},\ \bibinfo {year}
  {1980})\ \bibinfo {edition} {3rd}\ ed.,\ pp.\ \bibinfo {pages} {158 --
  190}\BibitemShut {NoStop}%
\end{thebibliography}%

\end{document}